\documentclass[onecolumn,eqsecnum,showpacs,amsmath,aps,pra]{revtex4}

 \usepackage{graphicx}
 \usepackage{bm}

\begin{document}

 \title{Medium-assisted vacuum force}
 \author{M. S. Toma\v s}
 \email{tomas@thphys.irb.hr}
 \affiliation{Rudjer Bo\v skovi\' c Institute, P. O. B. 180,
 10002 Zagreb, Croatia}
 \date{\today}

 \begin{abstract}
We discuss some implications of a very recently obtained result
for the force on a slab in a planar cavity based on the
calculation of the vacuum Lorentz force [C. Raabe and D.-G.
Welsch, Phys. Rev. A {\bf 71} (2005) 013814]. We demonstrate that,
according to this formula, the total force on the slab consists of
a medium-screened Casimir force and, in addition to it, a
medium-assisted force. The sign of of the medium-assisted force is
determined solely by the properties of the cavity mirrors. In the
Lifshitz configuration, this force is proportional to $1/d$ at
small distances and is very small compared with the corresponding
van der Waals force. At large distances, however, it is
proportional to $1/d^4$ and comparable with the Casimir force,
especially for denser media. The exponents in these power laws
decrease by 1 in the case of a thin slab. The formula for the
medium-assisted force also describes the force on a layer of the
cavity medium, which has similar properties. For dilute media, it
implies an atom-mirror interaction of the Coulomb type at small
and of the Casimir-Polder type at large atom-mirror distances. For
a perfectly reflecting mirror, the latter force is effectively
only three-times smaller than the Casimir-Polder force.


\end{abstract}
 \preprint{IRB-TH-2/05}
 \maketitle

\section{Introduction}
A number of approaches to the Casimir effect \cite{Cas} in
material systems lead to the conclusion that the Casimir force on
the medium between two bodies (mirrors) vanishes and that the only
existing force is that between the mirrors \cite{Schw,Zhou,Tom02}
(see also text books \cite{Abr,Mil} and references therein). It is
well known, however, that an atom (or a molecule) in the vicinity
of a mirror experiences the Casimir-Polder force \cite{CP} and, at
smaller distances, its nonretarded counterpart the van der Waals
force. Consequently, being a collection of atoms, every piece of a
medium in front of a mirror should experience the corresponding
force. To resolve this puzzling situation and overcome the above
"unphysical" result, usually derived by calculating the Minkowski
stress tensor \cite{Schw,Tom02} but also obtained using other
methods \cite{Schw,Zhou,Abr,Mil}, Raabe and Welsch \cite{Raa04}
very recently suggested an approach based on the calculation of
the vacuum Lorentz force (see also Ref. \cite{Obu}). In this
approach the force on a body is simply the sum of the Lorentz
forces acting on its constituents. Evidently, this should lead to
a nonzero force on the medium between the mirrors.

As an application of their approach, Raabe and Welsch calculated
the force on a magnetodielectric slab in a magnetodielectric
planar cavity. The aim of this work is to demonstrate several
straightforward implications of their formula. The paper is
organized as follows. For completeness, in Sec. II we (re)derive
the Raabe and Welsch formula and demonstrate that, according to
it, the force on the slab naturally splits into two rather
different components: a medium-screened and a medium-assisted
force. The latter force, being genuinely related to the
Lorentz-force approach, is discussed in more detail in Sec. III.
Our conclusions are summarized in Sec. IV. The necessary
mathematical background is given in the Appendices.

\section{Preliminaries}
Consider a multilayered system described by permittivity
 $\varepsilon({\bf r},\omega)=\varepsilon'({\bf r},\omega)+
 i\varepsilon''({\bf r},\omega)$ and permeability $\mu({\bf
 r},\omega)=\mu'({\bf r},\omega)+i\mu''({\bf r},\omega)$
 defined in a  stepwise fashion, as depicted in Fig. 1.
\begin{figure}[htb]
 \begin{center}
 \resizebox{10cm}{!}{\includegraphics{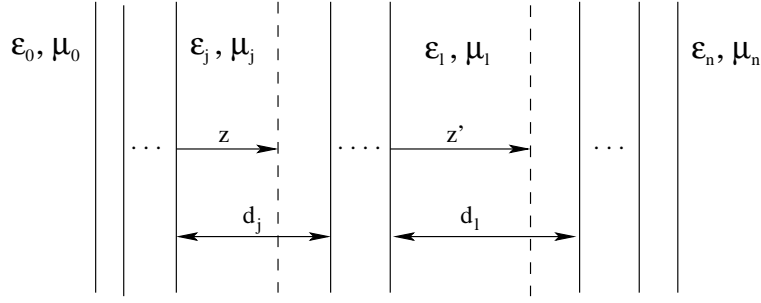}}
 \end{center}
 \caption{System considered schematically. The dashed lines
 represent the planes where the stress tensor is calculated.}
 \end{figure}
\noindent
 The force per unit area acting on a stack of layers between a plane
 $z$ in a $j$th layer and a plane $z'$ in an $l>j$ layer is then given by
 \begin{equation}
 \label{fjl}
 f_{jl}(z,z')=\tilde{T}_{l,zz}(z')-\tilde{T}_{j,zz}(z),
 \end{equation}
 where $\tilde{\tensor{\bf T}}_j\equiv \tensor{\bf T}_j-\tensor{\bf T}^0_j$,
 with $\tensor{\bf T}_j$ being the corresponding stress tensor
 and $\tensor{\bf T}^0_j$ its infinite-medium counterpart.

\subsection{Stress tensor}
The Lorentz-force approach to the Casimir effect eventually leads
to the calculation of the stress tensor (component)
\cite{Raa04,Obu}
\begin{equation}
 T_{j,zz}(z)=\frac{1}{8\pi}\left<E_zE_z-
 {\bf E}_\parallel\cdot{\bf E}_\parallel+B_zB_z-
 {\bf B}_\parallel\cdot{\bf B}_\parallel\right>_{{\bf r}\in(j)},
 \label{T}
 \end{equation}
where the brackets denote the average over the vacuum state of the
field. The correlation functions that appear here can be
straightforwardly calculated using the fluctuation-dissipation
theorem \cite{Aga,LiPi}. Decomposing the field operators into the
positive frequency and negative frequency parts according to
 \begin{equation}
 {\bf E}({\bf r},t)=\int_0^\infty d\omega
 {\bf E}({\bf r},\omega)e^{-i\omega t}+\int_0^\infty d\omega
 {\bf E}^\dag({\bf r},\omega)e^{i\omega t},
 \label{E}
 \end{equation}
we have (in the dyadic form) \cite{Aga}
 \begin{equation}
 \left<{\bf E}({\bf r},\omega){\bf E}^\dagger({\bf r}',\omega')\right>
 =\frac{\hbar}{\pi}\frac{\omega^2}{c^2}
 {\rm Im}\tensor{\bf G}({\bf r},{\bf r'};\omega)\delta(\omega-\omega'),
 \label{EE}
 \end{equation}
 and the magnetic-field correlation function is obtained from this
 expression using ${\bf B}({\bf r},\omega)=
 (-ic/\omega)\nabla\times{\bf E}({\bf r},\omega)$. Here
 $\tensor{\bf G}({\bf r},{\bf r'};\omega)$ is the classical Green
 function satisfying
 \begin{equation}
 \left[\nabla\times\frac{1}{\mu({\bf r},\omega)}\nabla\times-
 \varepsilon({\bf r},\omega)
 \frac{\omega^2}{c^2}\tensor{\bf I}\cdot\right]
 \tensor{\bf G}({\bf r},{\bf r'};\omega)=4\pi\tensor{\bf I}
 \delta({\bf r}-{\bf r'}),
 \label{GF}
 \end{equation}
 with the outgoing wave condition at the infinity.
Applying these results to the $j$th layer, for the relevant
 correlation functions we find
 \begin{subequations}
 \label{CF}
 \begin{equation}
 \left<{\bf E}({\bf r},t){\bf E}({\bf r},t)\right>_{{\bf r}\in(j)}
 =\frac{\hbar}{\pi}{\rm Im}\int_0^\infty d\omega
 \frac{\omega^2}{c^2}\tensor{\bf G}_j({\bf r},{\bf r};\omega),
 \label{ED}
 \end{equation}
 \begin{equation}
 \left<{\bf B}({\bf r},t){\bf B}({\bf r},t)\right>_{{\bf r}\in(j)}
 =\frac{\hbar}{\pi}{\rm Im}\int_0^\infty d\omega
 \tensor{\bf G}^B_j({\bf r},{\bf r};\omega),
 \label{BH}
 \end{equation}
 \end{subequations}
 where $\tensor{\bf G}_j({\bf r},{\bf r'};\omega)$ is the Green
 function element for ${\bf r}$ and ${\bf r'}$ both in the layer
 $j$, and
 \begin{equation}
 \tensor{\bf G}^B_j({\bf r},{\bf r'};\omega)=
 \nabla\times\tensor{\bf G}_j({\bf r},{\bf r'};\omega)\times
 \stackrel{\leftarrow}{\nabla'}
 \label{GB}
 \end{equation}
 is the corresponding Green function element for the magnetic field.

 With the above equations inserted in Eq. (\ref{T}), the stress
 tensor $\tilde{T}_{j,zz}$ is formally obtained by replacing the
 Green function with its scattering part
 \begin{equation}
 \tensor{\bf G}^{\rm sc}_j({\bf r},{\bf r'};\omega)= \tensor{\bf
 G}_j({\bf r},{\bf r'};\omega)- \tensor{\bf G}^0_j({\bf r},{\bf
 r'};\omega), \label{Gsc}
 \end{equation}
 where $\tensor{\bf G}^0_j({\bf r},{\bf r'};\omega)$ is the
 infinite-medium Green function. In this way, from Eq. (\ref{T})
 we have
 \begin{equation}
 \label{tT}
\tilde{T}_{j,zz}(z)=\frac{\hbar}{4\pi}{\rm Im}\int_0^\infty
 \frac{d\omega}{2\pi}\left\{\frac{\omega^2}{c^2}
 \left[G^{sc}_{j,zz}({\bf r},{\bf r};\omega)-
 G^{sc}_{j,\parallel}({\bf r},{\bf r};\omega)\right]
 +G^{B,sc}_{j,zz}({\bf r},{\bf r};\omega)-
 G^{B, sc}_{j,\parallel}({\bf r},{\bf r};\omega\right\},
 \end{equation}
 where $G^{sc}_{j,\parallel}({\bf r},{\bf r}';\omega)=
 G^{sc}_{j,xx}({\bf r},{\bf r}';\omega)+
 G^{sc}_{j,yy}({\bf r},{\bf r}';\omega)$.
In Appendix A, we derive the Green function
 $\tensor{\bf G}^{sc}_j({\bf r},{\bf r'};\omega)$ for a
 magnetodielectric multilayer and, in Appendix B, calculate
 the expression in the curly brackets of the above equation. We
 find that
 \begin{equation}
\label{cb}
\{\ldots\}=-2\pi i\mu_j\int\frac{d^2{\bf k}}{(2\pi)^2}
\frac{1}{\beta_j} \sum_{q=p,s}g_{qj}(\omega,k;z),
\end{equation}
where ${\bf k}$ and
$\beta_j(\omega,k)=\sqrt{n^2_j(\omega)\omega^2/c^2-k^2}$, with
$n_j(\omega)=\sqrt{\varepsilon_j(\omega)\mu_j(\omega)}$, are,
respectively, the parallel and the perpendicular component of the
wave vector in the layer, and the functions $g_{qj}(\omega,k;z)$
are in the shifted-z representation (see Appendix A) given by
\begin{eqnarray}
\label{gj}
g_{qj}(\omega,k;z)&=&\frac{2r^q_{j-}r^q_{j+}e^{2i\beta_j
d_j}}{D_{qj}}
\left[\beta^2_j(1+n^{-2}_j)+\Delta_qk^2(1-n^{-2}_j)\right]\nonumber\\
&+&\Delta_q\frac{r^q_{j-}e^{2i\beta_j z}+r^q_{j+}e^{2i\beta_j(d_j-
z)}} {D_{qj}}(\beta^2_j+k^2)(1-n^{-2}_j),\;\;\;\;0\leq z\leq d_j.
\end{eqnarray}
Here $\Delta_q=\delta_{qp}-\delta_{qs}$,
\begin{equation}
 D_{qj}(\omega,k)=1-r^q_{j-}r^q_{j+}e^{2i\beta_j d_j},
 \label{Dj}
 \end{equation}
 and $r^q_{j\pm}(\omega,k)$ are the reflection coefficients of
 the right and left stack bounding the layer, respectively.
 Specially, noting that $r^q_{0-}=r^q_{n+}=0$ and recalling that
 $d_0=0$ (see Appendix A), for the outmost (semi-infinite) layers
 we have
\begin{subequations}
\label{g0n}
\begin{eqnarray}
g_{q0}(\omega,k;z)&=&\Delta_q r^q_{0+}e^{-2i\beta_0
z}(\beta^2_0+k^2)
(1-n^{-2}_0),\;\;\;\;-\infty <z\leq 0,\\
g_{qn}(\omega,k;z)&=&\Delta_q r^q_{n-}e^{2i\beta_n
z}(\beta^2_n+k^2) (1-n^{-2}_n),\;\;\;\;0\leq z<\infty.
\end{eqnarray}
\end{subequations}

Converting the integral over the real $\omega$-axis in Eq.
(\ref{tT}) to that along the imaginary $\omega$-axis in the usual
way, letting $\omega=i\xi$,
 \begin{equation}
 \beta_j(i\xi,k)\equiv i\kappa_j(\xi,k)=
 i\sqrt{n^2_j(i\xi)\frac{\xi^2}{c^2}+k^2},
 \end{equation}
 and noticing the reality of the integrand, we finally obtain for
 the stress tensor in the layer \cite{Raa04}
 \begin{equation}
 \label{tTf}
 \tilde{T}_{j,zz}(z)=-\frac{\hbar}{8\pi^2}\int_0^\infty d\xi \mu_j
 \int^\infty_0\frac{dkk}{\kappa_j}\sum_{q=p,s}g_{qj}(i\xi,k;z).
 \end{equation}
As seen, the standard expression for the (Minkowski) stress tensor
obtained with \cite{Tom04}
 \begin{equation}
\label{gjM} g^M_{qj}(i\xi,k;z)=-4\kappa^2_j
\frac{r^q_{j-}r^q_{j+}e^{-2\kappa_j d_j}}{D_{qj}}
\end{equation}
is recovered from the above result only in the case of the empty
space between the stacks, i.e., only if
$\varepsilon_j(\omega)=\mu_j(\omega)=1$. We also note that,
according to Eq. (\ref{g0n}), the stress tensor is discontinuous
across the boundary between two semi-infinite media (in this case,
$0$ and $n$). This implies the existence of a force acting on a
layer around the interface between the media [$f_{\rm int}\equiv
f_{0n}(-a_0,a_n)$]
\begin{equation}
 f_{\rm int}=-\frac{\hbar}{8\pi^2c^2}\int_0^\infty d\xi\xi^2
 \int^\infty_0dkk\left[\frac{\mu_0}{\kappa_0}(n_0^2-1)e^{-2\kappa_0a_0}
 +\frac{\mu_n}{\kappa_n}(n_n^2-1)e^{-2\kappa_na_n}\right]
 \sum_{q=p,s}\Delta_qr^q_{0n}(i\xi,k;z),
 \label{fint}
 \end{equation}
where $a_0+a_n$ is the layer thickness and where we have used
$r^q_{0+}=-r^q_{n-}=r^q_{0n}$ [Eq. (\ref{rij})]. Since
$\tilde{T}^M_{zz}=0$ in semi-infinite layers, as follows from Eq.
(\ref{gjM}), such a force does not appear in the approach based on
the calculation of the Minkowski stress tensor \cite{com} and in
other equivalent approaches leading to the Lifshitz-like
expression \cite{Lif} for the force.

\subsection{Force in a planar cavity}

Owing to the $z$-dependence of $\tilde{T}_{j,zz}(z)$, Eqs.
(\ref{gj}) and (\ref{tTf}) imply the nonzero force on a slice of
the medium between the stacks contrary to the Lifshitz-like result
[Eqs. (\ref{tTf}) and (\ref{gjM})] obtained previously by many
authors \cite{Schw,Zhou,Tom02,Abr,Mil}. In order to calculate this
force, we consider a slightly more general configuration
consisting of a slab with refraction index $n_s$ and thickness
$d_s$ embedded in a material cavity with refraction index $n$ and
length $L$, as depicted in Fig. 2. The cavity walls are
conveniently described by the reflection coefficients $r^q_1$ and
$r^q_2$.
\begin{figure}[htb]
 \begin{center}
 \resizebox{8cm}{!}{\includegraphics{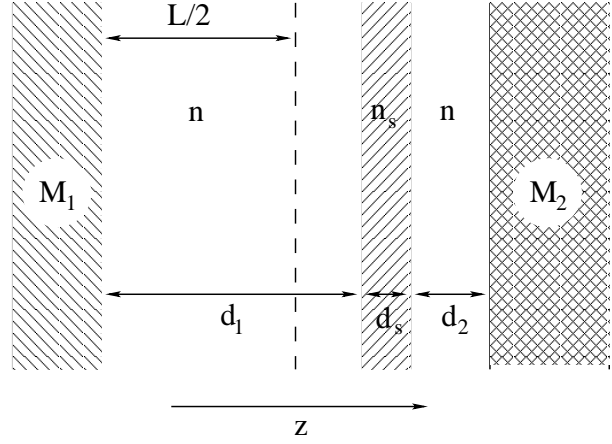}}
 \end{center}
 \caption{A slab in a planar cavity shown schematically. The
 refraction index of the slab is
 $n_s(\omega)=\sqrt{\varepsilon_s(\omega)\mu_s(\omega)}$ and that
 of the cavity $n(\omega)=\sqrt{\varepsilon(\omega)\mu(\omega)}$.
 The cavity walls are described by their reflection coefficients
 $r^q_1(\omega,k)$ and $r^q_2(\omega,k)$, with $k$ being the
 in-plane wave vector of a wave. The arrow indicates the direction
 of the force on the slab.}
\end{figure}

According to Eqs. (\ref{fjl}) and (\ref{tTf}), the force on the
slab $f_s=\tilde{T}_{2,zz}(0)-\tilde{T}_{1,zz}(d_1)$ in this
configuration is given by
\begin{equation}
\label{fs}
f_s(d_1,d_2)=-\frac{\hbar}{8\pi^2}\int_0^\infty d\xi \int^\infty_0
 dkk\frac{\mu}{\kappa}
\sum_{q=p,s}\left[g_{q2}(i\xi,k;0)-g_{q1}(i\xi,k;d_1)\right].
 \end{equation}
The functions $D_{q1}$ and $D_{q2}$ [Eq. (\ref{Dj})] are
straightforwardly obtained using Eq. (\ref{rijk}) to determine the
reflection coefficients at the right boundary of region 1
($r^q_{1+}$) and the left boundary of region 2 ($r^q_{2-}$). With
$r^q_{1-}=r^q_1$ and $r^q_{2+}=r^q_2$, we find
\begin{equation}
D_{q1}=1-r^q_1\left(r^q+\frac{{t^q}^2r^q_2 e^{2i\beta d_2}}
{1-r^qr^q_2 e^{2i\beta d_2}}\right)e^{2i\beta d_1}
\;\;\mathrm{and}\;\; D_{q2}=1-\left(r^q+\frac{{t^q}^2r^q_1
e^{2i\beta d_1}} {1-r^qr^q_1 e^{2i\beta d_1}}\right)r^q_2
e^{2i\beta d_2}.
\end{equation}
Here $r^q=r^q_{1/2}=r^q_{2/1}$ and $t^q=t^q_{1/2}=t^q_{2/1}$ are
Fresnel coefficients for the (whole) slab which are related
through [Eq. (\ref{rijk})]
\begin{equation}
\label{rt} r^q=\rho^q\frac{1-e^{2i\beta_sd_s}}
{1-{\rho^q}^2e^{2i\beta_s d_s}},\;\;\;
t^q=\frac{(1-{\rho^q}^2)e^{i\beta_s d_s}}{1-{\rho^q}^2e^{2i\beta_s
d_s}}
\end{equation}
to the {\it single-interface} medium-slab Fresnel reflection
coefficient $\rho^q=r^q_{1s}=r^q_{2s}$, given by [see Eq.
(\ref{rij})]
\begin{equation}
\rho^q=\frac{\beta-\gamma^q\beta_s} {\beta+\gamma^q\beta_s},\;\;\;
\gamma^p=\frac{\varepsilon}{\varepsilon_s},\;\;
\gamma^s=\frac{\mu}{\mu_s}.
\label{rho}
\end{equation}
This gives
\begin{eqnarray}
\label{g21}
g_{q2}(\omega,k;0)-g_{q1}(\omega,k;d_1)
&=&\left\{4\beta^2\left(\delta_{qs}+\frac{1}{n^2}\delta_{qp}\right)r^q
+\frac{\omega^2}{c^2}(n^2-1)[(1+r^q)^2-{t^q}^2] \Delta_q\right\}
\nonumber\\
&&\times\frac{r^q_2e^{2i\beta d_{2}}-r^q_1e^{2i\beta d_{1}}}{N^q},
\end{eqnarray}
where
\begin{equation}
N^q =1-r^q (r^q _1e^{2i\beta d_{1}}+r^q _2e^{2i\beta d_2})
+({r^q }^2-{t^q }^2)r^q _1r^q _2e^{2i\beta (d_1+d_2)}.
\end{equation}

Combining Eqs. (\ref{fs}) and (\ref{g21}), we see that $f_s$
naturally splits into two rather different components
\begin{equation}
\label{fsf} f_s(d_1,d_2)=f^{(1)}(d_1,d_2)+f^{(2)}(d_1,d_2),
\end{equation}
where
\begin{equation}
\label{f1} f^{(1)}(d_1,d_2)=\frac{\hbar}{2\pi^2}\int_0^\infty d\xi
\int^\infty_0 dkk\kappa\sum_{q=p,s}\left(\mu\delta_{qs}+
\frac{1}{\varepsilon}\delta_{qp}\right) r^q\frac{r^q_2e^{-2\kappa
d_{2}}-r^q_1e^{-2\kappa d_{1}}}{N^q},
\end{equation}
and
\begin{equation}
\label{f2}
 f^{(2)}(d_1,d_2)=\frac{\hbar}{8\pi^2c^2}\int_0^\infty
 d\xi\xi^2\mu(n^2-1)\int^\infty_0\frac{dkk}{\kappa}
 \sum_{q=p,s}[(1+r^q)^2-{t^q}^2]\Delta_q
\frac{r^q_2e^{-2\kappa d_{2}}-r^q_1e^{-2\kappa d_{1}}}{N^q}.
 \end{equation}
Equation (\ref{f1}) differs in two respects from the formula for
the Casimir force in a dielectric cavity obtained through the
Minkowski tensor calculation \cite{Tom02}. First, the Fresnel
coefficients refer to a magnetodielectric system \cite{Tom04}.
Another new feature in Eq. (\ref{f1}) is the (effective) screening
of the force through the multiplication of the contributions
coming from TE- and TM-polarized waves by $\mu$ and
$1/\varepsilon$, respectively. This gives a simple recipe how to
adapt the traditionally obtained formulas for the Casimir force to
the Lorentz-force approach, as we illustrate below.

Clearly, $f^{(2)}_s$ owes its appearance to the cavity medium
(note that it vanishes when $n=1$) and is therefore a genuine
consequence of the Lorentz force approach, so that below we
consider this force in more detail.

\section{Medium-assisted force}

\subsection{Force on a slab}
Assuming, for simplicity, a large (semi-infinite) cavity obtained
formally by letting $d_1\rightarrow\infty$ (or $r^q_1=0$), from Eq.
(\ref{f2}), we have
\begin{equation}
\label{f2i}
 f^{(2)}(d)=\frac{\hbar}{8\pi^2c^2}\int_0^\infty
 d\xi\xi^2\mu(n^2-1)\int^\infty_0\frac{dkk}{\kappa}
 \sum_{q=p,s}\Delta_q\frac{[(1+r^q)^2-{t^q }^2]
R^qe^{-2\kappa d}}{1-r^qR^qe^{-2\kappa d}},
 \end{equation}
where we have changed the notation so that $d_2\equiv d$ and
$r^q_2\equiv R^q$. Another remarkable feature of the
medium-assisted force is that its sign depends only on the
properties of the mirror. Indeed, assuming an ideally reflecting
mirror and letting $R^q=\pm\Delta_q$ (the minus sign is for an
infinitely permeable mirror, see Eq. (\ref{Rps}) below), we
clearly see that $ f^{(2)}$ is attractive or repulsive, depending
on whether the mirror is (dominantly) conducting (dielectric) or
permeable irrespective of the properties of the slab.

\subsubsection{Small distances}

The integral over $\xi$ in Eq. (\ref{f2i}) effectively extends up
to a frequency $\Omega$ beyond which the mirror becomes
transparent. Accordingly, at small mirror-slab distances
$d\ll\Lambda=2\pi c/\Omega$ the main contribution to $f^{(2)}$
comes from large $k$'s ($k\sim 1/d$). In this region, the
nonretarded (quasistatic) approximation applies to the integrand
obtained formally by letting $\kappa=\kappa_l=k$ everywhere. Thus,
for example, for a structureless mirror consisting of a
semi-infinite medium with refraction index $n_m$ we have [from Eq.
(\ref{rij})]
\begin{equation}
 R^p_{{\rm nr}\infty}(i\xi,k)=\frac{\varepsilon_m-\varepsilon}
 {\varepsilon_m+\varepsilon}\equiv \rho(\varepsilon_m,\varepsilon),
 \;\;\; R^s_{{\rm nr}\infty}(i\xi,k)=\frac{\mu_m-\mu}{\mu_m+\mu},\label{Rqs}
\end{equation}
and the nonretarded Fresnel coefficients of the slab are from Eq.
(\ref{rt}) given by
\begin{equation}
\label{rqs} r^q_{\rm nr}(i\xi,k)=\rho^q_{\rm nr}
\frac{1-e^{-2kd_s}}{1-[\rho^q_{\rm nr}]^2e^{-2kd_s}}, \;\;\;\;
t^q_{\rm nr}(i\xi,k)=\frac{(1-\rho^q_{\rm nr})^2e^{-kd_s}}
{1-[\rho^q_{\rm nr}]^2e^{-2kd_s}},
\end{equation}
with $\rho^p_{\rm nr}=\rho(\varepsilon_s,\varepsilon)$ and
$\rho^s_{\rm nr}=\rho(\mu_s,\mu)$ [see Eq. (\ref{rho})]. With the
substitution $u=2kd$, this gives
\begin{equation}
\label{f2s}
 f^{(2)}(d\ll \Lambda)=\frac{\hbar}{16\pi^2c^2d}\int_0^\infty
 d\xi\xi^2\mu(n^2-1)\int^\infty_0du
 \sum_{q=p,s}\Delta_q\frac{[(1+r^q)^2-{t^q}^2]_{\rm nr}
R^q_{\rm nr}e^{-u}}{1-r^q_{\rm nr}R^q_{\rm nr}e^{-u}},
 \end{equation}
where the (nonretarded) reflection coefficients are now functions
of $(i\xi,\frac{u}{2d})$.

The medium-assisted force on a thick, $d_s\rightarrow\infty$, slab
at small distances is obtained from the above equation when
letting $t^q_{\rm nr}=0$ and $r^q_{\rm nr}=\rho^q_{\rm nr}$ [see
Eq. (\ref{rqs})]. Specially, in the case of a single-medium mirror,
corresponding to the classical Lifshitz (L) configuration
\cite{Lif}, all reflection coefficients in Eq. (\ref{f2s}) are
independent of $u$ so that the entire dependence of $f^{(2)}$ on
$d$ is given by the factor in front of the integral. Using Eq.
(\ref{Rqs}), in this case we find
\begin{eqnarray}
\label{f2sL} f^{(2)}_{\rm L}(d\ll \Lambda;d_s\gg d)&=&
\frac{\hbar}{16\pi^2c^2d}\int_0^\infty
d\xi\xi^2\mu(n^2-1)\int^\infty_0
du\left\{\left(\frac{2\varepsilon_s}{\varepsilon_s+\varepsilon}\right)^2
\left[\frac{\varepsilon_m+\varepsilon}
{\varepsilon_m-\varepsilon}e^u-\frac{\varepsilon_s-\varepsilon}
{\varepsilon_s+\varepsilon}\right]^{-1} \right.\nonumber\\
&&- \left. \left(\frac{2\mu_s}{\mu_s+\mu}\right)^2
\left[\frac{\mu_m+\mu} {\mu_m-\mu}e^u-\frac{\mu_s-\mu}
{\mu_s+\mu}\right]^{-1} \right\}.
\end{eqnarray}
We compare this with the screened Casimir force in the Lifshitz
configuration which, by applying the recipe embodied in Eq.
(\ref{f1}) directly to the Lifshitz formula \cite{Lif}, reads
\begin{equation}
\label{fL_qs} f^{(1)}_{\rm L}(d\ll \Lambda;d_s\gg d)=
\frac{\hbar}{16\pi^2d^3}\int_0^\infty d\xi
\int^\infty_0duu^2\left\{
\frac{1}{\varepsilon}\left[\frac{\varepsilon_s+\varepsilon}
{\varepsilon_s-\varepsilon}\frac{\varepsilon_m+\varepsilon}
{\varepsilon_m-\varepsilon}e^u-1\right]^{-1}+\mu
\left[\frac{\mu_s+\mu}{\mu_s-\mu}
\frac{\mu_m+\mu}{\mu_m-\mu}e^u-1\right]^{-1}\right\}.
\end{equation}
If we scale the frequency in the above integrals with $\Omega$, we
see that $f^{(2)}_{\rm L}/f^{(1)}_{\rm L}\sim (\Omega d/c)^2\ll 1$
Accordingly, the medium-assisted force at small distances is very
small when compared with the screened van der Waals force.

Of interest is also the medium-assisted force on a thin, $d_s\ll
d$, slab. From Eqs. (\ref{rt}) and (\ref{rho}) we find that to the
first order in $\kappa_sd_s$
\begin{equation}
\label{ths}
 r^q(i\xi,k)\simeq 2\rho^q\kappa_s d_s,\;\;\;\;
[(1+r^q)^2-{t^q}^2](i\xi,k)\simeq 2\frac{\kappa d_s}{\gamma^q}.
\end{equation}
Making here the nonretarded approximation ($\kappa_s=\kappa=k$)
and letting $k\rightarrow u/2d$, from Eq. (\ref{f2s}) we find that
to the first order in $d_s/d$
\begin{equation}
\label{f2st}
 f^{(2)}(d\ll \Lambda;d_s\ll d)=\frac{\hbar d_s}{16\pi^2c^2d^2}\int_0^\infty
 d\xi\xi^2\mu(n^2-1)\int^\infty_0duue^{-u}\left[\frac{\varepsilon_s}
{\varepsilon}R^p_{\rm nr}(i\xi,\frac{u}{2d})
-\frac{\mu_s}{\mu}R^s_{\rm nr}(i\xi,\frac{u}{2d})\right],
 \end{equation}
which, for a single-medium (s-m) mirror, reduces to
\begin{equation}
\label{f2st2}
 f^{(2)}_{\rm s-m}(d\ll \Lambda;d_s\ll d)=\frac{\hbar d_s}{16\pi^2c^2d^2}\int_0^\infty
 d\xi\xi^2\mu(n^2-1)\left(\frac{\varepsilon_s}
{\varepsilon}\frac{\varepsilon_m-\varepsilon}
{\varepsilon_m+\varepsilon}
-\frac{\mu_s}{\mu}\frac{\mu_m-\mu}{\mu_m+\mu} \right).
 \end{equation}

\subsubsection{Large distances}

To find $f^{(2)}$ for large $d$, we use the standard substitution
$\kappa=n\xi p/c$ in Eq. (\ref{f2i}). This gives
\begin{equation}
\label{f2l}
 f^{(2)}(d)=\frac{\hbar}{8\pi^2c^3}\int_0^\infty
 d\xi\xi^3\mu n(n^2-1)\int^\infty_1dp
 \sum_{q=p,s}\Delta_q\frac{[(1+r^q)^2-{t^q}^2]
R^qe^{-2n\xi pd/c}}{1-r^qR^qe^{-2n\xi pd/c}},
 \end{equation}
where the reflection coefficients as functions of $(i\xi,p)$ are
obtained from their $(i\xi,k)$-counterparts by letting
\begin{equation}\label{sl}
\kappa_l\rightarrow n\frac{\xi}{c}s_l,\;\;\;
s_l=\sqrt{p^2-1+n^2_l/n^2}
\end{equation}
for all relevant layers. Thus, for example, for a single-medium
mirror we have [from Eq. (\ref{rij})]
\begin{equation}
\label{Rps} R^p_\infty(i\xi,p)= \frac{\varepsilon_m p-\varepsilon s_m}
{\varepsilon_m p+\varepsilon s_m}\equiv\rho(\varepsilon_m,\varepsilon;p),\;\;\;
R^s_\infty(i\xi,p)=\frac{\mu_m p-\mu s_m}{\mu_m p+\mu s_m}.
\end{equation}
Now, since $p\geq 1$, for large $d$ the contributions
from the $\xi\simeq 0$ region dominate the integral in Eq.
(\ref{f2l}). Consequently, we may approximate the
frequency-dependent quantities with their static values (which we
denote by the subscript $0$). With the substitution $v=2n_0\xi
pd/c$, this leads to
\begin{equation}
\label{f2l1}
 f^{(2)}(d\gg \Lambda)=\frac{\hbar c\mu_0(n_0^2-1)}{2^7\pi^2n_0^3d^4}\int_0^\infty
 dvv^3\int^\infty_1\frac{dp}{p^4}
 \sum_{q=p,s}\Delta_q\frac{[(1+r^q)^2-{t^q}^2]_0
R^q_0e^{-v}}{1-r^q_0R^q_0e^{-v}}.
 \end{equation}

For the Lifshitz configuration
[$t^q=0$, $r^p=\rho(\varepsilon_s,\varepsilon;p)$, $r^s=\rho(\mu_s,\mu;p)$ and
$R^q=R^q_\infty $, see Eq. (\ref{Rps})], we now obtain
\begin{eqnarray}
\label{f2lL} f^{(2)}_{\rm L}(d\gg \Lambda;d_s\gg d)&=&\frac{\hbar
c\mu_0(n_0^2-1)}{2^7\pi^2n_0^3d^4}\int_0^\infty
 dvv^3\int^\infty_1\frac{dp}{p^4}
\left\{\left(\frac{2\varepsilon_s p}{\varepsilon_s p+\varepsilon s_s}\right)_0^2
\left[\frac{\varepsilon_m p+\varepsilon s_m}
{\varepsilon_m p-\varepsilon s_m}e^v-\frac{\varepsilon_s p-\varepsilon s_s}
{\varepsilon_s p+\varepsilon s_s}\right]_0^{-1} \right.\nonumber\\
&&\left.-\left(\frac{2\mu_s p}{\mu_s p+\mu s_s}\right)_0^2
\left[\frac{\mu_m p+\mu s_m} {\mu_m p-\mu s_m}e^v-\frac{\mu_s
p-\mu s_s} {\mu_s p+\mu s_s}\right]_0^{-1}\right\},
\end{eqnarray}
which is to be compared with the screened Casimir force at large
distances \cite{Lif}
\begin{eqnarray}
\label{f1lL} f^{(1)}_{\rm L} (d\gg \Lambda;d_s\gg d)&=&\frac{\hbar
c}{2^5\pi^2n_0d^4}\int_0^\infty
 dvv^3\int^\infty_1\frac{dp}{p^2}\left\{\frac{1}{\varepsilon_0}
\left[\frac{\varepsilon_{s}p+\varepsilon s_{s}}
{\varepsilon_{s}p-\varepsilon
s_{s}}\frac{\varepsilon_{m}p+\varepsilon s_{m}}
{\varepsilon_{m}p-\varepsilon s_{m}}e^v-1\right]_0^{-1} \right.\nonumber\\
&&\left.+\mu_0 \left[\frac{\mu_{s}p+\mu s_{s}} {\mu_{s}p-\mu
s_{s}}\frac{\mu_{m}p+\mu s_{m}} {\mu_{m}p-\mu
s_{m}}e^v-1\right]_0^{-1}\right\}.
\end{eqnarray}
The relative magnitude of $f^{(2)}$ and $f^{(1)}$ is best
estimated if we consider the force in a cavity with ideally
reflecting mirrors, corresponding to the classical Casimir
configuration. Letting $\varepsilon_{s0}\rightarrow\infty$ and
$\varepsilon_{m0}\rightarrow\infty$, the integrals in Eqs.
(\ref{f2lL}) and (\ref{f1lL}) become elementary and we find
\begin{equation}
f^{(2)}_{\rm id}(d\gg \Lambda)=\frac{\hbar c\pi^2}{45\cdot
2^5d^4}\sqrt{\frac{\mu_0}{\varepsilon_0}}
\left(1-\frac{1}{n_0^2}\right),
\end{equation}
\begin{equation}
f^{(1)}_{\rm id}(d\gg \Lambda)=\frac{\hbar c\pi^2}{15\cdot
2^5d^4}\sqrt{\frac{\mu_0}{\varepsilon_0}}
\left(1+\frac{1}{n_0^2}\right),
\end{equation}
It is seen that at large distances $f^{(2)}$ is comparable in
magnitude with $f^{(1)}$, especially for optically denser media
where, ideally, $f^{(2)}$ is only three times smaller than
$f^{(1)}$.

To find the force on a thin slab at large distances , we note that
according to Eq. (\ref{ths})
\begin{equation}
r^q(i\xi,p)=2\rho^q\frac{n\xi s_sd_s}{c},\;\;\;
[(1+r^q)^2-{t^q}^2](i\xi,p)\simeq 2\frac{n\xi pd_s} {c\gamma^q}.
\label{ths2}
\end{equation}
Inserting this into Eq. (\ref{f2l}) and proceeding in the same way
as above, we find to the first order in $d_s/d$
\begin{equation}
\label{f2lt}
 f^{(2)}(d\gg \Lambda;d_s\ll d))=\frac{3\hbar c\mu_0(n_0^2-1)d_s}{16\pi^2n_0^3d^5}
 \int^\infty_1\frac{dp}{p^4}
 \left[\frac{\varepsilon_{s0}}{\varepsilon_0}R^p(0,p)-
 \frac{\mu_{s0}}{\mu_0}R^s(0,p)\right].
 \end{equation}

\subsection{Force on the cavity medium}
Clearly, when $n_s=n$, $f^{(2)}_s$ describes the force on a layer
of the medium in the cavity $f_m$. Since in this case $\rho^q=0$
in Eq. (\ref{rt}), the corresponding results for $f_m$ are
straightforwardly obtained from the above formulas when letting
$r^q(i\xi,k)=0$ and $t^q(i\xi,k)=e^{-\kappa d_s}$. Thus, from Eq.
(\ref{f2i}) we find that $f_m$ is generally given by
\begin{equation}
\label{fmg}
 f_m(d)=\frac{\hbar}{8\pi^2c^2}\int_0^\infty
 d\xi\xi^2\mu(n^2-1)\int^\infty_0\frac{dkk}{\kappa}(1-e^{-2\kappa d_s})e^{-2\kappa d}
 \sum_{q=p,s}\Delta_qR^q(i\xi,k).
 \end{equation}
The small-distance behavior of $f_m$ from Eq. (\ref{f2s}) is described by
\begin{equation}
\label{fms}
 f_m(d\ll \Lambda)=\frac{\hbar}{16\pi^2c^2d}\int_0^\infty
 d\xi\xi^2\mu(n^2-1)\int^\infty_0du(1-e^{-ud_s/d})e^{-u}
 \sum_{q=p,s}\Delta_qR^q_{\rm nr}(i\xi,\frac{u}{2d}),
 \end{equation}
and, as follows from Eq. (\ref{f2l1}) (upon performing the
integration over $v$), at large distances $f_m$ behaves as
\begin{equation}
\label{fml}
 f_m(d\gg \Lambda)=\frac{3\hbar c\mu_0(n_0^2-1)}{64\pi^2n_0^3}
\left[\frac{1}{d^4}-\frac{1}{(d+d_s)^4}\right]\int^\infty_1\frac{dp}{p^4}
 \sum_{q=p,s}\Delta_qR^q(0,p).
\end{equation}
Note that for an ideally reflecting mirror the value of the above
integral is $\pm 2/3$. Accordingly, the force on the medium is
attractive or repulsive depending on whether the mirror is
(dominantly) dielectric or permeable resembling, in this respect,
the force on an (electrically polarizable) atom
\cite{Boy1,Buh1,Buh2} near a mirror.

The thick-layer results are easily recognized from the above
formulas when letting $d_s\gg d$. Similarly, the force on a thin
layer is given by these equations in the limit $d_s\ll d$. At
small distances, from Eq. (\ref{fms}) we find
\begin{subequations}
\label{fmst}
\begin{eqnarray}
f_m(d\ll \Lambda;d_s\ll d)&=&\frac{\hbar
d_s}{16\pi^2c^2d^2}\int_0^\infty
 d\xi\xi^2\mu(n^2-1)\int^\infty_0duue^{-u}
 \sum_{q=p,s}\Delta_qR^q_{\rm nr}(i\xi,\frac{u}{2d}),\\
 &=&\frac{\hbar d_s}{16\pi^2c^2d^2}\int_0^\infty
 d\xi\xi^2\mu(n^2-1)\left(\frac{\varepsilon_m-\varepsilon}{\varepsilon_m+\varepsilon}-
 \frac{\mu_m-\mu} {\mu_m+\mu}\right)
 \end{eqnarray}
 \end{subequations}
in agreement with Eq. (\ref{f2st}). Here the second line
corresponds to the system with a structureless mirror.  Finally,
the force on a thin layer at large distances is from Eq.
(\ref{fml}) found to be
\begin{equation}
\label{fmlt}
 f_m(d\gg \Lambda;d_s\ll d)=\frac{3\hbar c(n_0^2-1)d_s}
 {16\pi^2n_0\varepsilon_0d^5}
\int^\infty_1\frac{dp}{p^4} \left[R^p(0,p)-R^s(0,p)\right],
\end{equation}
in agreement with Eq. (\ref{f2lt}).

We end this short discussion by noting that for a dilute medium
$f_m$ is the sum of the forces $f_{ai}$ acting on each atom $i$ in
the layer. Accordingly, the force on an atom $f_a$ at distance $d$
from a mirror is obtained from $f_m$ for a thin layer as
$f_a=f_m/Nd_s$, where $N$ is the atomic number density. Since for
dilute media $n^2-1=4\pi N(\alpha_e+\alpha_m)$, it follows that
$f_a$ is given by the above thin-layer results upon making the
formal replacement
\begin{equation}
\frac{n^2(i\xi)-1}{4\pi}d_s\rightarrow\alpha_e(i\xi)+\alpha_m(i\xi),
\end{equation}
where $\alpha_{e(m)}$ is the electric (magnetic) polarizability of
the atom. Thus, expanding the integrand in Eq. (\ref{fmg}) for
small $2\kappa d_s\sim d_s/d$ and using the above recipe, we find
that generally
\begin{equation}
\label{fmgt}
 f_a(d)=\frac{\hbar}{\pi c^2}\int_0^\infty
 d\xi\xi^2\mu(\alpha_e+\alpha_m)\int^\infty_0 dkk e^{-2\kappa d}
 \left[R^p(i\xi,k)-R^s(i\xi,k)\right].
 \end{equation}
We also observe that Eq. (\ref{fmst}) then implies a Coulomb-like
force on an atom at small distances from a mirror rather than the
common van der Waals force \cite{Zhou}. At large atom-mirror
distances, however, Eq. (\ref{fmlt}) implies a screened
Casimir-Polder force on the atom. Of course, in accordance with
the above mentioned unique property of the medium-assisted force,
the sign of $f_a$ is insensitive to the polarizability type
(electric or magnetic) of the atom contrary to the standard
Casimir-Polder force \cite{Boy2}. Note also that, since
$n_0\varepsilon_0\simeq 1$ for dilute media, $f_a$ at large
distances from an ideally reflecting dielectric mirror is
effectively three times smaller than the Casimir-Polder force. We
stress, however, that, as a medium-assisted force, $f_a$ is a
collective property of the atomic system and this (perhaps)
explains its unusual properties.

It is natural to compare the above medium-assisted atomic force
with the familiar force $\tilde{f}_a$ acting on an atom in vacuum
near a mirror. This {\it single-atom} force can be obtained in the
same way as above by considering the force on a thin dilute slab
in an empty semi-infinite cavity. We find
\begin{equation}
\label{tfa}
 \tilde{f}_a(d)=\frac{\hbar}{\pi c^2}\int_0^\infty
 d\xi\xi^2\int^\infty_0 dkk e^{-2\kappa d}\left\{
 \left[\alpha_e\left(2\frac{\kappa^2c^2}{\xi^2}-1\right)-
 \alpha_m\right]R^p(i\xi,k)
+\left[\alpha_m\left(2\frac{\kappa^2c^2}{\xi^2}-1\right)-
\alpha_e\right]R^s(i\xi,k)\right\},
 \end{equation}
which generalizes (in different directions) earlier results
obtained for $\tilde{f}_a$ in various systems
\cite{Schw,Zhou,CP,Boy1,Buh1,Buh2,Boy2}. This expression correctly
reproduces the dependence of the Casimir- Polder force on the
polarizability type of the atom \cite{Boy2} and the
dielectric/magnetic properties of the mirror
\cite{Boy1,Buh1,Buh2}. Also, for structureless mirrors,
$\tilde{f}_a\sim 1/d^4$ at small and $\tilde{f}_a\sim 1/d^5$ at
large distances. Apparently, this asymptotic behaviour of the
atom-mirror force is well supported experimentally
\cite{Ori,San,Suk,Lan,Shi,Dru,Lin,Pas}. However, we note that the
results presented in these works do not definitely disqualify the
medium-assisted force. Indeed, being a collective property, $f_a$
is expected to show up at higher atomic densities, whereas most
experiments were usually performed with low-density atomic beams
\cite{San,Suk,Lan,Shi,Dru,Pas}, i.e. under the conditions in
favour of the single-atom force. Besides, a number of these
experiments probed the $d^{-5}$ tail of the force
\cite{Suk,San,Shi,Lin,Pas}, which is common to both $f_a$ and
$\tilde{f_a}$. Actually, there were also spectroscopic evidences
showing that the characteristic features due to the $d^{-4}$ tail
of $\tilde{f_a}$ disappear from the spectra at higher atomic
densities \cite{Ori}. Accordingly, to test the existence of $f_a$,
one must design an experiment involving a higher-density
homogeneous atomic system close to a mirror and probing the
nonretarded atom-mirror interaction, where $f_a$ substantially
differs from $\tilde{f}_a$. On the theoretical side, to understand
the properties of the medium-assisted force, a microscopic
consideration of the atom-mirror interaction is needed, for an
atom of the medium in the vicinity of a mirror.

\section{Summary}
In summary, in this work we have discussed a formula for the force
on a slab in a planar cavity, as derived very recently by Raabe
and Welsch using the Lorentz-force approach \cite{Raa04}. We have
shown that this result naturally splits into a formula for a
medium-screened Casimir force and a formula for a medium-assisted
force. A remarkable feature of the latter force is that its sign
depends only on the properties of the cavity mirrors. In the
classical Lifshitz configuration, at small distances the
medium-assisted force is proportional to $d^{-1}$ and is generally
very small compared with the screened van der Waals force ($\sim
d^{-3})$. At large distances, however, the medium-assisted force
is proportional to $d^{-4}$ and is comparable with the screened
Casimir force, especially for denser media (actually, for a dense
medium in a cavity with ideally reflecting mirrors, it is only
three times smaller). As usual, the exponents in these power laws
decrease by 1 in the case of a thin slab. The formula for the
medium-assisted force also describes the force on the cavity
medium. For dilute media, it predicts the atom-mirror interaction
of the Coulomb type at small and of the Casimir-Polder type at
large atom-mirror distances. In a semi-infinite cavity with an
ideally reflecting mirror, the predicted medium-assisted force on
an atom is effectively only three times smaller at large distances
than the Casimir-Polder force.

\acknowledgments{This work was supported by the Ministry of
Science and Technology of the Republic of Croatia under contract
No. 0098001.}

 \appendix
 \section{Green function}

 Following the derivation presented in Ref. \cite{Tom95}
 for a purely dielectric multilayer, for clarity,
 we consider the field
 \begin{equation}
 {\bf E}({\bf r},{\bf r}';\omega) =\frac{\omega^2}{c^2}
 \tensor{\bf G}({\bf r},{\bf r}';\omega)\cdot{\bf p}
 \label{EdG}
 \end{equation}
 of an oscillating point dipole ${\bf p}\exp{(-i\omega t)}$ at a
 position ${\bf r}'$ rather than the Green function itself.
 Assuming the dipole in a $j$th layer, its field ${\bf
 E}^{(j)}_l({\bf r},{\bf r}';\omega)$ in an $l$th layer is given
 by
 \begin{equation}
 {\bf E}^{(j)}_l({\bf r},{\bf r}';\omega)=
 {\bf E}^0_j({\bf r},{\bf r}';\omega)\delta_{lj}+
 {\bf E}^h_l({\bf r},{\bf r}';\omega),
 \end{equation}
 where ${\bf E}^0_j({\bf r},{\bf r}';\omega)$ is the field of
 the dipole as would be in the infinite medium $j$ and
 ${\bf E}^h_l({\bf r},{\bf r}';\omega)$ describe the
 propagation of this source field through the system.
 Specially, ${\bf E}^h_j({\bf r},{\bf r}';\omega)\equiv
 {\bf E}^{sc}_j({\bf r},{\bf r}';\omega)$ represents
 the scattered (reflected) field in the $j$th layer.

 According to Eq. (\ref{GF}), ${\bf E}^0_j({\bf r},{\bf r}';\omega)$
 is of the same form as the dipole field in a purely dielectric
 medium multiplied by $\mu_j$ except that this time the wave
 vector is given by $k_j=n_j\omega/c=\sqrt{\varepsilon_j\mu_j}\omega/c$.
 In the plane-wave representation
 \begin{equation}
 {\bf E}({\bf r},{\bf r}';\omega)=\int\frac{d^2{\bf k}}{(2\pi)^2}
 {\bf E}({\bf k},\omega;z,z')e^{i{\bf k}\cdot({\bf r}_\parallel-
 {\bf r}'_\parallel)},
 \label{Ekz}
 \end{equation}
 we therefore have \cite{Tom95}
 \begin{equation}
 {\bf E}^0_j({\bf k},\omega;z,z')=-4\pi\frac{\mu_j}{\varepsilon_j}
 \hat{\bf z}\hat{\bf z}\cdot{\bf p}\delta(z-z')
 +\sum_{q=p,s}\left[\hat{\bf e}^+_{qj}({\bf k})e^{i\beta_jz}E^{0+}_{qj}
 \theta(z-z')+\hat{\bf e}^-_{qj}({\bf k})e^{-i\beta_jz}E^{0-}_{qj}
 \theta(z'-z)\right],
 \label{E0j}
 \end{equation}
 where $\beta_j=\sqrt{k_j^2-k^2}$,
 \begin{equation}
 E^{0\pm}_{qj}=\mu_j\frac{2\pi i}{\beta_j}\frac{\omega^2}{c^2}\xi_q
 \hat{\bf e}^\mp_{qj}(-{\bf k})\cdot{\bf p}\;e^{\mp i\beta_jz'},
 \label{E0qj}
 \end{equation}
 with $\xi_q=\delta_{qp}-\delta_{qs}$, and
 \begin{equation}
 \hat{\bf e}^\pm_{pj}({\bf k})=\frac{1}{k_j}(k\hat{\bf z}\mp\beta_j
 \hat{\bf k}),\;\;\;\hat{\bf e}^\pm_{sj}({\bf k})= \hat{\bf
 k}\times\hat{\bf z}\equiv\hat{\bf n},
 \end{equation}
 are unit polarization vectors for $q=p$ (TM) and $q=s$ (TE)
 polarized waves, respectively.

 The fields ${\bf E}^h_l({\bf r},{\bf r}';\omega)$ obey homogeneous
 Maxwell equations. In analogy to Eq. (\ref{E0j}), ${\bf
 E}^h_l({\bf k},\omega;z,z')$ can therefore be written as
 \begin{equation}
 {\bf E}^h_l({\bf k},\omega;z,z')=\sum_{q=p,s}
 \left[\hat{\bf e}^+_{ql}({\bf k})e^{i\beta_lz}E^+_{ql}
 +\hat{\bf e}^-_{ql}({\bf k})e^{-i\beta_lz}E^-_{ql}\right].
 \label{Ehl}
 \end{equation}
 Since only the outgoing waves should exist in the external layers,
 $E^+_{q0}=E^-_{qn}=0$ and the remaining coefficients $E^\pm_{ql}$
 can be expressed in terms of the generalized reflection and
 transmission coefficients of the corresponding stacks of layers. A
 reflection coefficient $r^q$ of a stack is defined as the ratio of
 the reflected to incoming wave (electric-field) amplitude (factors
 multiplying $\hat{\bf e}$'s) at the corresponding stack's
 boundary. Similarly, a transmission coefficient $t^q$ of a stack
 is defined as the ratio of the transmitted to incident wave
 amplitude calculated at the corresponding stack's boundaries. In
 calculating these coefficients it is convenient to adopt a
 (shifted-z) representation for the field \cite{Tom95} in which
 $0\leq z\leq d_l$ in any finite layer, whereas $-\infty<z\leq 0$ ($l=0$) and
 $0\leq z<\infty$ ($l=n$), respectively, in the external layers.

 According to the above definitions, the coefficients $E^\pm_{qj}$
 of the field in the $j$th layer are given by
\begin{equation}
 E^+_{qj}=r^q_{j-}(E^{0-}_{qj}+E^-_{qj}),\;\;\;\;\;
e^{-i\beta_jd_j}E^-_{qj}=r^q_{j+}e^{i\beta_jd_j}(E^{0+}_{qj}+E^+_{qj}),
 \end{equation}
where we have introduced the notation $r^q_{j-}\equiv r^q_{j/0}$
 and $r^q_{j+}\equiv r^q_{j/n}$ for the reflection coefficients of
 the bounding stacks. With Eq. (\ref{E0qj}), we find
 \begin{subequations}
 \label{Eqj}
 \begin{equation}
 E^+_{qj}=\mu_j\frac{2\pi i}{\beta_j}\frac{\omega^2}{c^2}\xi_q
 \frac{r^q_{j-}e^{i\beta_jd_j}}{D_{qj}}
 [\hat{\bf e}^+_{qj}(-{\bf k})e^{-i\beta_jz'_+}
 +r^q_{j+}\hat{\bf e}^-_{qj}(-{\bf k})e^{i\beta_jz'_+}]\cdot{\bf p},
\end{equation}
 \begin{equation}
 E^-_{qj}=\mu_j\frac{2\pi i}{\beta_j}\frac{\omega^2}{c^2}\xi_q
 \frac{r^q_{j+}e^{2i\beta_jd_j}}{D_{qj}}
 [\hat{\bf e}^-_{qj}(-{\bf k})e^{-i\beta_jz'_-}
 +r^q_{j-}\hat{\bf e}^+_{qj}(-{\bf k})e^{i\beta_jz'_-}]\cdot{\bf p},
 \end{equation}
 \end{subequations}
 where $z'_+\equiv d_j-z'$ and $z'_-\equiv z'$ are the
 distances of the dipole from the layer's boundaries and
 \begin{equation}
 D_{qj}=1-r^q_{j-} r^q_{j+} e^{2i\beta_j d_j}.
 \end{equation}
 Repeating the same considerations for the dipole embedded in the
 layer $0$ ($n$), we find that its field
 ${\bf E}^{sc}_0({\bf r},{\bf r}';\omega)$
 [${\bf E}^{sc}_n({\bf r},{\bf r}';\omega)$] is also given by the
 above equations, with $j=0$ $(n)$, provided that we let
 $r^q_{0-}=0$ ($r^q_{n+}=0$) and put $d_0$ $(d_n)$, which appears formally
 in Eq. (\ref{Eqj}), equal to zero.

 Collecting the equations and using Eq. (\ref{EdG}), we obtain the
 Green function for the scattered field in the $j$th layer in the
 form
 \begin{widetext}
 \begin{eqnarray}
 \tensor{\bf G}^{\rm sc}_j({\bf r},{\bf r}';\omega)&=&
 \mu_j\frac{i}{2\pi}\int\frac{d^2{\bf k}}
 {\beta_j}e^{i{\bf k}\cdot({\bf r}_\parallel-{\bf r}_\parallel')}
 \sum_{q=p,s}\xi_q\frac{e^{i\beta_j d_j}}{D_{qj}}
 \left\{r^q_{j-}\hat{\bf e}^+_{qj}({\bf k})e^{i\beta_j z_-}
 \left[\hat{\bf e}^+_{qj}(-{\bf k})e^{-i\beta_jz'_+}+
 r^q_{j+}\hat{\bf e}^-_{qj}(-{\bf k})e^{i\beta_jz'_+}\right]\right.
 \nonumber\\
 &&\left.+r^q_{j+}\hat{\bf e}^-_{qj}({\bf k})e^{i\beta_j z_+}
 \left[\hat{\bf e}^-_{qj}(-{\bf k})e^{-i\beta_jz'_-}+
 r^q_{j-}\hat{\bf e}^+_{qj}(-{\bf
 k})e^{i\beta_jz'_-}\right]\right\},\;\;\;\;\;0\leq z,z'\leq d_j.
  \label{greensc}
 \end{eqnarray}
 \end{widetext}
 Apparently, except for the multiplication by
 $\mu_j$, $\tensor{\bf G}^{\rm sc}_j({\bf r},{\bf r}';\omega)$ is
 formally the same as for a purely dielectric system.
 This time, however, the wave vectors in the layers are given
 by $k_l=\sqrt{\varepsilon_l\mu_l}\omega/c$. As follows from their
 definition, for local stratified media the Fresnel coefficients
 satisfy recurrence and symmetry relations
 \begin{subequations}
 \label{rrel}
 \begin{equation}
 r^q_{i/j/k}=r^q_{i/j}+\frac{t^q_{i/j}t^q_{j/i}r^q_{j/k}
 e^{2i\beta_jd_j}}{1-r^q_{j/i}r^q_{j/k}e^{2i\beta_jd_j}},
 \label{rijk}
 \end{equation}
 \begin{equation}
 t^q_{i/j/k}=\frac{t^q_{i/j}t^q_{j/k}e^{i\beta_jd_j}}
 {1-r^q_{j/i}r^q_{j/k}e^{2i\beta_jd_j}}=
 \frac{\mu_k\beta_i}{\mu_i\beta_k}t^q_{k/j/i},
 \label{tijk}
 \end{equation}
 \end{subequations}
 and, for a single $i-j$ interface, reduce to
 \begin{subequations}
 \label{sic}
 \begin{equation}
 \label{rij}
 r^q_{ij}=\frac{\beta_i-\gamma^q_{ij}\beta_j}
 {\beta_i+\gamma^q_{ij}\beta_j}=-r^q_{ji},
 \end{equation}
 \begin{equation}
 \label{tij}
 t^q_{ij}=\sqrt{\frac{\gamma^q_{ij}}{\gamma^s_{ij}}}(1+r^q_{ij})=
 \frac{\mu_j\beta_i}{\mu_i\beta_j}t^q_{ji},
 \end{equation}
 \end{subequations}
 where $\gamma^p_{ij}=\varepsilon_i/\varepsilon_j$ and
 $\gamma^s_{ij}=\mu_i/\mu_j$.

 \section{Calculation of Eq. (\ref{cb})}
 Performing the derivations indicated in Eq. \ (\ref{GB}) and using
\begin{equation}
{\bf K}^\pm_j({\bf k})\times\hat{\bf e}^\pm_{qj}({\bf k})=
k_j\xi_q\hat{\bf e}^\pm_{q'j}({\bf k}),\;\;\;p'=s,\;\;s'=p,
\end{equation}
we find that $\tensor{\bf G}^{B,\rm sc}_j({\bf r},{\bf r}';\omega)$
is given by Eq. (\ref{greensc}) multiplied by $-k_j^2$ and
with $\hat{\bf e}^\pm_{qj}\rightarrow\hat{\bf e}^\pm_{q'j}$.
Noting that the equal-point Green function dyadics consist only of
diagonal elements, we easily find
\begin{equation}
\tensor{\bf G}_j^{\rm sc}({\bf r},{\bf r};\omega)=
\frac{i\mu_j}{2\pi k_j^2}\int\frac{d^2{\bf k}}{\beta_j}
\left\{\hat{\bf k}\hat{\bf k}\frac{\beta_j^2}{D_{pj}}
\left[2r^p_{j-}r^p_{j+}e^{2i\beta_j d_j}-r^p_{j-}e^{2i\beta z_-}
-r^p_{j+}e^{2i\beta z_+}\right]\right.
\end{equation}
\begin{equation}
\left.+\hat{\bf n}\hat{\bf n}\frac{k_j^2}{D_{sj}}
\left[2r^s_{j-}r^s_{j+}e^{2i\beta_j d_j}+r^s_{j-}e^{2i\beta_j z_-}
+r^s_{j+}e^{2i\beta_j z_+}\right]+\hat{\bf z}\hat{\bf
z}\frac{k^2}{D_{pj}} \left[2r^p_{j-}r^p_{j+}e^{2i\beta_j
d_j}+r^p_{j-}e^{2i\beta_j z_-} +r^p_{j+}e^{2i\beta_j
z_+}\right]\right\},\nonumber
\end{equation}
and $\tensor{\bf G}^{B,\rm sc}_j({\bf r},{\bf r};\omega)$ is given
by this equation multiplied by $k_j^2$ and with
$p\leftrightarrow s$. The traces $G^{\rm sc}_{j,\parallel}({\bf
r},{\bf r};\omega)$ and $G^{B,\rm sc}_{j,\parallel}({\bf r},{\bf
r};\omega)$ can be easily recognized from these equations and one
has, for example,
\begin{equation}
\frac{\omega^2}{c^2}\left[G^{\rm sc}_{j,zz}({\bf r},{\bf
r};\omega) -G^{\rm sc}_{j,\parallel}({\bf r},{\bf
r};\omega)\right]= \frac{i\mu_j}{2\pi n^2_j}\int\frac{d^2{\bf
k}}{\beta_j}\left\{\frac{k^2}{D_{pj}}
\left[2r^p_{j-}r^p_{j+}e^{2i\beta_j d_j}+r^p_{j-}e^{2i\beta_j z_-}
+r^p_{j+}e^{2i\beta_j z_+}\right]\right.
\end{equation}
\begin{equation}
\left.-\frac{\beta_j^2}{D_{pj}}
\left[2r^p_{j-}r^p_{j+}e^{2i\beta_j d_j}-r^p_{j-}e^{2i\beta z_-}
-r^p_{j+}e^{2i\beta z_+}\right]\nonumber\\
-\frac{k_j^2}{D_{sj}} \left[2r^s_{j-}r^s_{j+}e^{2i\beta_j
d_j}+r^s_{j-}e^{2i\beta_j z_-} +r^s_{j+}e^{2i\beta_j
z_+}\right]\right\},\nonumber
\end{equation}
while $G^{B,\rm sc}_{j,zz}({\bf r},{\bf r};\omega)- G^{B,\rm
sc}_{j,\parallel}({\bf r},{\bf r};\omega)$ is given by this
equation multiplied by $n^2_j$ and with $p\leftrightarrow s$.
Adding these two quantities, one obtains Eq. (\ref{cb}) for the
expression in the curly bracket of Eq. (\ref{tT}).


\begin{thebibliography}{00}

\bibitem{Cas}H. B. G. Casimir, Proc. K. NED. Akad.
  Wet. {\bf 51} (1948) 793.

\bibitem{Schw}J. Schwinger, L. L. DeRaad, Jr., and K. A. Milton,
Ann. Phys. (N. Y.) {\bf 115}, 1 (1978).

\bibitem{Zhou}F. Zhou and L. Spruch, Phys. Rev. A {\bf 52}
(1995) 297.

\bibitem{Tom02}M. S. Toma\v s, Phys. Rev. A {\bf 66} (2002)
052103.


\bibitem{Abr} A. A. Abrikosov, L. P. Gorkov, and I. E. Dzyaloshinski,
{\it Methods of Quantum Field Theory in Statistical Physics},
(Prentice-Hall, Englewood Cliffs, NJ, 1963) Ch 6.

\bibitem{Mil}P. W. Milonni, {\it The Quantum Vacuum.
An Introduction to Quantum Electrodynamics} (San Diego,
Academic Press, 1994) Chap. 7.

\bibitem{CP}H. B. G. Casimir and D. Polder, Phys. Rev. {\bf 73}
(1948) 360.

\bibitem{Raa04}C. Raabe and D.-G. Welsch,
Phys. Rev. A {\bf 71}, (2005) 013814.

\bibitem{Obu}   Y. N. Obukhov and F. W. Hehl, Phys. Lett. A
{\bf 311} (2003) 277.

\bibitem{Aga}G. S. Agarwal, Phys. Rev. A, {\bf 11} (1975) 230.

\bibitem{LiPi} E. M. Lifshitz and L. P. Pitaevskii,
{\it Statistical Physics}, Part 2, (Pergamon Press, Oxford, 1991)
Ch 8.

\bibitem{Tom04}M. S. Toma\v s, e-print arXiv: quant-ph/0410057.

\bibitem{com}The interface force found in Ref. \cite{Schw}
(and dropped as unobservable) was obtained from the volume
contribution to the Minkowski stress tensor.

\bibitem{Lif} E. M. Lifshitz, Zh. Eksp. Teor. Fiz.
{\bf 29} (1955) 94  [Sov. Phys. JETP {\bf 2} (1956) 73 ].

\bibitem{Boy1}T. H. Boyer, Phys. Rev. A {\bf 9}, (1974) 2078.

\bibitem{Buh1}S. Y. Buhmann, H. T. Dung, T. Kampf, and D.-G.
Welsch, e-print arXiv: quant-ph/0501168.

\bibitem{Buh2} S. Y. Buhmann and D.-G. Welsch, e-print arXiv:
quant-ph/0502183.

\bibitem{Boy2}T. H. Boyer, Phys. Rev. {\bf 180}, (1969) 19.

\bibitem{Ori}M. Oria, M. Chevrollier, D. Bloch, M. Fichet, and
M. Ducloy, Europhys. Lett. {\bf 14} (1991) 527.

\bibitem{San}V. Sandoghdar, C. I. Sukenik, and E. A. Hinds,
Phys. Rev. Lett. {\bf 68}, (1992) 3432.

\bibitem{Suk}C. I. Sukenik, M. G. Boshier, D. Cho, V. Sandoghdar, and E. A.
Hinds, Phys. Rev. Lett. {\bf 70}, (1993) 560.

\bibitem{Lan}A. Landragin, J.-Y. Courtois, G. Labeyrie, N.
Vansteenkiste, C. I. Westbrook, and A. Aspect, Phys. Rev. Lett.
{\bf 77}, (1996) 1664.

\bibitem{Shi}F. Shimizu, Phys. Rev. Lett. {\bf 86}, (2001) 987.

\bibitem{Dru}V. Druzhinina and M. DeKieviet,
Phys. Rev. Lett. {\bf 91}, (2003) 193202.

\bibitem{Lin}Y. Lin, I. Teper, C. Chin, and V. Vuleti\' c,
Phys. Rev. Lett. {\bf 92}, (2004) 050404.

\bibitem{Pas}T. A. Pasquini, Y. Shin, C. Sanner, M. Saba, A.
Schirotzek, D. E. Pritchard, and W. Ketterle, Phys. Rev. Lett.
{\bf 93}, (2004) 223201.

\bibitem{Tom95}M. S. Toma\v s, Phys. Rev. A {\bf 51} (1995) 2545.


 \end{thebibliography}
 \end{document}